\documentclass[aps,twocolumn,pra,superscriptaddress,showpacs,tightenlines]{revtex4-1}
\usepackage{amsmath}
\usepackage{graphicx}
\usepackage{epsfig}
\usepackage{amsfonts}
\usepackage{color}
\usepackage{epstopdf}

\begin{document}

\newcommand{\nn}{\nonumber}
\newcommand{\ms}[1]{\mbox{\scriptsize #1}}
\newcommand{\msi}[1]{\mbox{\scriptsize\textit{#1}}}
\newcommand{\dg}{^\dagger}
\newcommand{\smallfrac}[2]{\mbox{$\frac{#1}{#2}$}}
\newcommand{\Tr}{\text{Tr}}
\newcommand{\ket}[1]{|#1\rangle}
\newcommand{\bra}[1]{\langle#1|}
\bibliographystyle{apsrev}
\newcommand{\pfpx}[2]{\frac{\partial #1}{\partial #2}}
\newcommand{\dfdx}[2]{\frac{d #1}{d #2}}
\newcommand{\half}{\smallfrac{1}{2}}
\newcommand{\s}{{\mathcal S}}
\newcommand{\jord}{\color{red}}
\newcommand{\kurt}{\color{blue}}

\title{Implementation of weak measurement amplification with the weak coherent light and optomechanical system}
\author{Gang Li}
\email{ligang0311@sina.cn}
\affiliation{Beijing Computational Science Research Center, Beijing, 100193, People's Republic of China}
\affiliation{School of Physics and Electronic Information, Yan'an University, Yanan, 716000, People's Republic of China}
\author{Wen Yang}
\email{wenyang@csrc.ac.cn}
\affiliation{Beijing Computational Science Research Center, Beijing, 100193, People's Republic of China}

\date{\today}

\begin{abstract}
Most studies for postselected weak measurement in optomechanical system focus on using a single photon as a measured system. However, we find that using weak coherent light instead of a single photon can also amplify the mirror's position displacement of one photon. In the WVA regime (Weak Value Amplification regime), the weak value of one photon can lie outside the eigenvalue spectrum, proportional to the difference of the mirror's position displacement between the successful and failed postselection and its successful postselection probability is dependent of the mean photon number of the coherent light and improved by adjusting it accordingly. Outside the WVA regime, the amplification limit can reach the level of the vacuum fluctuations, and when the mean photon number and the optomechanical coupling parameter are selected appropriately, its successful postselection probability becomes higher, which is beneficial to observe the maximum amplification value under the current experimental conditions. This result breaks the constraint that it is difficult to detect outside the WVA regime. This opens up a new regime for the study of a single photon nonlinearity in optomechanical system.
\end{abstract}

\maketitle

\section{ Introduction}

Weak measurement theory, first introduced by Aharonov, Albert and Vaidman\cite{Aharonov88}, describes a measurement situation where the measurementof a property of a system is usually performed by coupling the system to a pointer, and the state of the pointer changes depending on the value of this property. In weak measurement (and postselection is performed)\cite{Aharonov88}, the pointer's displacement, averaged over many measurementrepetitions, is proportional to the `weak value' of the observable: $\langle A\rangle_{W}=\langle\psi_{f}|A|\psi_{i}
\rangle/\langle\psi_{f}|\psi_{i}\rangle$, where $A$ is the observable that describes a property of the system, and $|\psi_{i}\rangle$ and $|\psi_{f}\rangle$ are the initial (preselected) and final (postselected) states of the system, respectively. Strangely, the weak value is not constrained to be within the eigenvalue spectrum of the observable $A$ and not even in general a real number\cite{Jozsa07}, but these results occur at the expense of reducing the postselection probability. Since the weak value is larger than any of the eigenvalues, there is an amplification effect on the pointer. In this case, it is called the Weak-Value Amplification regime (WVA regime).

Weak measurements and weak values can be used to study fundamental issues in quantum mechanics and explain certain paradoxes, such as wave-particle duality\cite{Aharonov17}, nonlocality\cite{Aharonov16-}, contextuality\cite{Aharonov-16}, quantum shell game\cite{Ravon07,Aharon08,Resch04},Cheshire cats\cite{Aharonov16,Denkmayr17}, macroscopic nonclassical state and Hardy's paradox\cite{Hardy92,Aharonov02,Lundeen09}. Its important characteristic is used to amplify tiny physical effects, such as ultrasensitive beam deflection\cite{Dixon09}, the transverse separation of an optical beam due to birefringence\cite{Ritchie91,Hosten08}, longitudinal
phase shifts\cite{Brunner10}, and angular rotations of a classical beam\cite{Magana-Loaiza14}, and to measure small physical quantities, such as the direct measurement of wave functions\cite{Lundeen11}, and observation of the average trajectories in a two slit interferometer\cite{Kocsis11}. Li et al.\cite{Li14,Li15,Li-2015,Li20,Carrasco19} showed that the mirror's displacement of a single photon in optomechanical system\cite{Girvin09,Marquardt13} can be amplified and observed using weak measurements when the initial states of the mirror are in the different quantum states. These schemes require single-photon sources, and the successful postselection probability is very low. However, the conditions for preparing an ideal single photon are very harsh and the economic cost is very high, so these schemes\cite{Li14,Carrasco19} are difficult to be verified by experiments. Feizpour et al.\cite{Feizpour11,Feizpour15,Hallaji17} showed that the nonlinear cross-Kerr effect of a single photon can be amplified using weak measurements. In\cite{Hallaji17} weak value amplification can amplify the nonlinear cross-Kerr effect of a single photon caused by a classical light and the successful postselection probability can be improved by adjusting the mean photons of coherent light accordingly. And using weak value based on classical light discuss the classical limit of the radiation pressure and the difference in interpretation arising from classical and quantum descriptions, by treating one of the mirrors of an interferometer quantum mechanically\cite{Aharonov13}. Their results motivate a possibility to detect the mirror's displacement of one photon by replacing single photons with coherent light in the postselection weak-coupling optomechanical system.

In this paper, we show that using weak coherent light instead of single photon can also amplify the mirror's position displacement caused by one photon. In the WVA regime, the weak value of one photon can lie outside the eigenvalue spectrum, proportional to the difference of the mirror's position displacement between the successful and failed postselection and its successful postselection probability is dependent of the mean photon number of the coherent light and improved by adjusting it accordingly. Hence, we were able to maintain a substantial set of postselections even for small the postselection parameter (i.e., the inner product of the preselected and the postselected states) simply by adjusting the mean photon number accordingly. Outside the WVA regime, the amplification limit can reach the level of the vacuum fluctuations, and when the mean photon number and the optomechanical coupling parameter range are selected appropriately, its successful postselection probability becomes higher, as compared to the studies of weak measurement with one photon and optomechanical system\cite{Li14,Carrasco19}, which is beneficial to observe the maximum amplification value under the current experimental conditions. This result is unexpected and it breaks the constraint that it is difficult to detect outside the WVA regime. Outside the WVA regime, the quantum-limited signal to noise ratio (quantum-limited SNR) is increasing, even reaching one.

The structure of our paper is as follows. In Sec. II, we give weak measurement model in optomechanical system, In Sec. III and IV, we state the
main results based on the weak coherent light in this work, including the amplification effect of the mirror's position displacement and weak value of one photon, and In Sec. V, we give the discussion and conclusion about the work.

\section{Optomechanical model}

Here we will consider a weak measurement model in optomechanical system where the initial state of the system (the signal light) is prepared in weak coherent state.
\begin{figure}[tbp]
\centering
\includegraphics[scale=0.5]{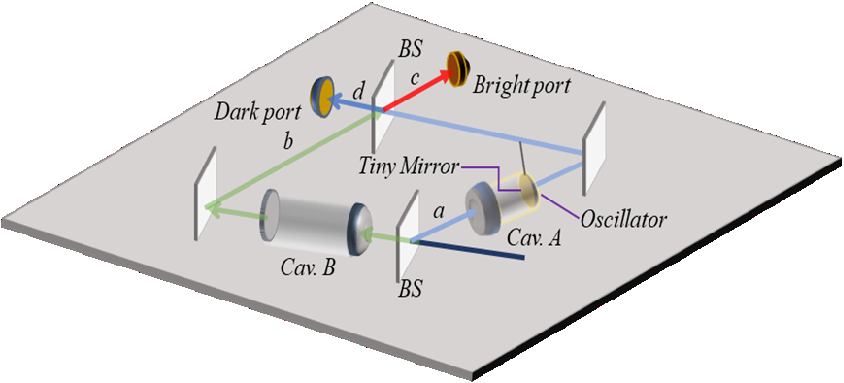}. \centering
\caption{The weak light enters the first beam splitter of March-Zehnder interferometer, before entering an optomechanical cavity A and a
conventional cavity B. The coherent light weakly excites the tiny mirror. After the second beam splitter, dark port detect one photon and bright port also detect the photons, i.e., postselection acts on the case where the mirror has been excited by an added photon and multiphoton, however, bright port can detect the photons and no photon be detected in dark port, i.e., postselection is fails.}
\end{figure}
Now consider a Mach-Zehnder interferometer shown in Fig. 1. The optomechanical cavity A is embedded in one arm of the March-Zehnder
interferometer and a stationary Fabry-Perot cavity B is placed in another arm. The first beam splitter is symmetric, but the second one is
unsymmetric. The Hamiltonian of the optomechanical system in the interferometer\cite{Bouwmeester12,Li14} is given by (set $\hbar=1$)
\begin{equation}
H=\omega_{0}(a_{a}^{\dagger}a_{a}+a_{b}^{\dagger}a_{b})+\omega_{m}c^{\dagger}c-ga_{a}^{\dagger}a_{a}(c^{\dagger}+c),  \label{a1}
\end{equation}
where $\omega_{0}$ is the frequency of the optic cavity A, B of length $L$ with corresponding annihilation operators $a_{a}$ and $a_{b}$, $\omega_{m}$ is the angular frequency of mechanicalsystem with corresponding annihilation operator $c$, and the optomechanical coupling strength $g=\omega_{0}\sigma/L$, $\sigma=(1/2M\omega_{m})^{1/2}$ which is the zero point fluctuation and $M$ is the effective mass of mechanical system. In this article, it is a weak measurement model where the mirror is used as the pointer to measure the photon nmuber $a_{a}^{\dagger}a_{a}$ in cavity A.

\section{ The expected position displacement of the mirror in optomechanics based on a weak coherent light}

As shown in Fig. 1, suppose a weak coherent light is input into the interferometer, the state of the light after the first beam splitter becomes $|\psi_{i}\rangle=|\alpha/\sqrt{2}\rangle_{a}|\alpha/\sqrt{2}\rangle_{b}$ (Preselected state). The initial state of the mirror is prepared at the ground state $|0\rangle_{m}$ (Gaussian state). After the weak interaction in Eq. (\ref{a1}), according to the results of the Hamiltonian in\cite{Mancini97,Bose97}, the combined state of the light and the mirror will be
\begin{eqnarray}
|\psi_{evo}\rangle&=&\exp [-ir\omega_{m}t(a_{a}^{\dagger}a_{a}+a_{b}^{\dagger}a_{b})]\exp[i\phi(a_{a}^{\dagger}a_{a})^{2}]  \nonumber \\
&&\exp[a_{a}^{\dagger}a_{a}(\varphi c^{\dagger}-{\varphi}^{\ast}c)]\exp[-i\omega_{m}tc^{\dagger}c]  \nonumber \\
&&|\alpha/\sqrt{2}\rangle_{a}|\alpha/\sqrt{2}\rangle_{b}|0\rangle_{m}, \label{a3}
\end{eqnarray}
where $r=\omega_{0}/\omega_{m}$, $\phi=k^{2}(\omega_{m}t-\sin\omega_{m}t)$, $\varphi=k(1-e^{-i\omega_{m}t})$, $k=g/\omega_{m}$ is the scaled
coupling parameter.

The second imbalanced beam splitter can be modeled via
\begin{equation}
\binom{a_{c}}{a_{d}}=\left(
\begin{array}{cc}
\cos\theta&\sin\theta \\ \sin\theta&-\cos\theta
\end{array}
\right)\binom{a_{a}}{a_{b}}  \label{a4}
\end{equation}

Therefore, the combined state after the second beam splitter can be written as,
\begin{eqnarray}
|\psi_{evo}\rangle&=&\exp[-ir\omega_{m}t(a_{c}^{\dagger}a_{c}+a_{d}^{\dagger}a_{d})]\exp[i\phi(\cos^{2}\theta a_{c}^{\dagger}a_{c}
\nonumber \\
&+&\cos\theta\sin\theta(a_{c}^{\dagger}a_{d}+a_{c}a_{d}^{\dagger})+\sin^{2}\theta a_{d}^{\dagger}a_{d})^{2}]  \nonumber \\
&&\exp[(\cos^{2}\theta a_{c}^{\dagger}a_{c}+\cos\theta\sin\theta(a_{c}^{\dagger}a_{d}+a_{c}a_{d}^{\dagger})  \nonumber \\
&+&\sin^{2}\theta a_{d}^{\dagger}a_{d})(\varphi c^{\dagger}-{\varphi}^{\ast}c)]\exp[-i\omega _{m}tc^{\dagger }c]  \nonumber \\
&&|\alpha(\cos\theta+\sin\theta)/\sqrt{2}\rangle_{c}|\alpha(\sin\theta   \nonumber \\
&-&\cos\theta)/\sqrt{2}\rangle_{d}|0\rangle_{m}  \label{a5}
\end{eqnarray}
Note that the initial state of the light is rewritten in the representations of the arm $c$ and the arm $d$ as $|\psi_{i}\rangle=|\alpha(\cos\theta+\sin\theta)/\sqrt{2}\rangle_{c}|\alpha(\sin\theta-\cos\theta)/\sqrt{2}\rangle_{d}$.

We take the limit where the overlap between the preselected and postselected states is very close to $1$, and port $d$ in Fig.1 is nearly dark. This implies that in Eq. (\ref{a4}), $\theta\approx\pi/4$. For $\theta=\pi/4+\delta $ ($\delta\ll1$, called the postselection parameter), we
expand $\cos\theta$ and $\sin\theta$ as $\cos\theta=(1-\delta)/\sqrt{2}$ and $\sin\theta=(1+\delta)/\sqrt{2}$, respectively. And because $|\varphi|=\sqrt{2}k(1-\cos\omega_{m}t)\ll1$, i.e., $k\ll1$. Therefore, the state in Eq. (\ref{a5}) becomes (see appendix A for detail derivation)
\begin{eqnarray}
|\psi_{evo}\rangle&=&\exp[\alpha e^{-ir\omega_{m}t}(a_{c}^{\dagger}+a_{d}^{\dagger})(\varphi c^{\dagger}  \nonumber \\
&-&{\varphi}^{\ast}c)/2]|\alpha e^{-ir\omega_{m}t}\rangle_{c}|\delta\alpha e^{-ir\omega_{m}t}\rangle_{d}|0\rangle_{m}.  \label{a9}
\end{eqnarray}

Next we only consider the process of weak measurement, i.e., mathematically, by adjusting the inner product of the preselection and the postselection states $\langle\psi_{f}|\psi_{i}\rangle$ and comparing it with the optomechanical coupling strength\cite{Li15,Li20}. However, when $\delta$ ($\langle\psi_{f}|\psi_{i}\rangle\approx\alpha\delta$) is much larger than the optomechanical coupling strength $k$, i.e., $\delta\gg k$, which is called the Weak-Value Amplification regime (WVA regime). In addition, the others are called outside the Weak-Value Amplification regime (outside the WVA regime). For the optomechanical device of Fig. 1, we record whether or not the detector fires, to measure the expected position displacement written on the mirror separately for the events of successful postselection (which we term `click') and the events of failed postselection (`no-click'). Here we consider weak signal light instead of single photon. For $|\alpha|^{2}\delta^{2}\ll1$, one can write this state $|\delta\alpha e^{-ir\omega_{m}t}\rangle_{d}$ in the arm $d$, the dark port, as $|\delta\alpha e^{-ir\omega_{m}t}\rangle _{d}\approx|0\rangle_{d}+\delta\alpha e^{-ir\omega_{m}t}|1\rangle_{d}$, where $|0\rangle_{d}$ and $|1\rangle_{d}$ are vacuum and single-photon number states, respectively.

When the single photon detector does not fire, we project onto $|0\rangle_{d}$, i.e., in the language of weak measurement the failed postselected state of the light in dark port is $|0\rangle_{d}$, then the state of the light and the mirror becomes
\begin{eqnarray}
|\psi_{no-click}\rangle&=&\langle0|_{d}\exp[\alpha e^{-ir\omega_{m}t}(a_{c}^{\dagger}+a_{d}^{\dagger})(\varphi c^{\dagger}  \nonumber \\
&-&{\varphi}^{\ast}c)/2]|\alpha e^{-ir\omega_{m}t}\rangle_{c}|\delta\alpha e^{-ir\omega_{m}t}\rangle_{d}|0\rangle_{m}  \nonumber \\
&=&\exp[\alpha e^{-ir\omega_{m}t}a_{c}^{\dagger}(\varphi c^{\dagger}-{\varphi}^{\ast }c)/2]  \nonumber \\
&&|\alpha e^{-ir\omega_{m}t}\rangle_{c}|0\rangle_{m}.  \label{a10}
\end{eqnarray}

Substituting (\ref{a10}) into (\ref{B1}) (see appendix B), in bright port, the expected position displacement on the mirror will be
\begin{equation}
\langle q\rangle_{no-click}=\sigma(\varphi+{\varphi}^{\ast})|\alpha|^{2}/2.  \label{a12}
\end{equation}
This result shows that for the coherent state, the expected position displacement of the mirror is $|\alpha|^{2}/2$ times $\sigma(\varphi+{\varphi}^{\ast})$, where $\sigma(\varphi+{\varphi}^{\ast})$ is the mirror's position displacement caused by a single photon (see appendix C). Therefore, the results above are caused by the classical light effect ($\propto|\alpha|^{2}/2$).

On the other hand, when the detector in the dark port fires, we project onto $|1\rangle_{d}$, i.e., in the language of weak measurement the successful postselected state of the light in dark port is $|1\rangle_{d}$, then the state of the light and the mirror becomes
\begin{eqnarray}
|\psi_{click}\rangle&=&P_{f}^{-1}\langle 1|_{d}\exp[\alpha e^{-ir\omega_{m}t}(a_{c}^{\dagger}+a_{d}^{\dagger})(\varphi c^{\dagger}
\nonumber \\
&-&{\varphi}^{\ast}c)/2]|\alpha e^{-ir\omega_{m}t}\rangle_{c}|\delta\alpha e^{-ir\omega_{m}t}\rangle_{d}|0\rangle_{m}  \nonumber \\
&=&P_{f}^{-1}\exp[\alpha e^{-ir\omega_{m}t}a_{c}^{\dagger}(\varphi c^{\dagger}-{\varphi}^{\ast}c)/2]  \nonumber \\
&&|\alpha e^{-ir\omega_{m}t}\rangle_{c}(\delta\alpha|0\rangle_{m}+\varphi\alpha/2|1\rangle_{m}),  \label{a13}
\end{eqnarray}
where $P_{f}=|\alpha|^{2}(\delta^{2}+|\varphi|^{2}/4)$ is the successful postselection probability.

Substituting (\ref{a13}) into (\ref{B1}), in dart port, the expected position displacement on the mirror is given by
\begin{equation}
\langle q\rangle_{click}=\sigma(\varphi+{\varphi}^{\ast})[|\alpha|^{2}/2+\delta/(2\delta^{2}+|\varphi|^{2}/2)].  \label{a15}
\end{equation}
This result also shows that, for the single photon state based on weak measurement, the expected position displacement of the mirror is $\delta/(2\delta^{2}+|\varphi|^{2}/2)$ times $\sigma(\varphi+{\varphi}^{\ast})$. Therefore, the above results are caused by the classical light effect ($\propto|\alpha|^{2}/2$) and the single photon nonlinear effect of weak measurement ($\propto\delta/(2\delta^{2}+|\varphi|^{2}/2)$)\cite{Li14}. Moreover, it can be seen from Eq. (\ref{a15}) that its maximal value is $\sigma(\varphi+\varphi^{\ast})(|\alpha|^{2}/2+1/2\sqrt{k(\varphi+{\varphi}^{\ast})})$\ when $\delta\alpha=|\varphi|\alpha /2$. For example, when $\omega_{m}t=\pi$, the maximal value of $\langle q\rangle_{click}$ is $\sigma(1+2k|\alpha|^{2})$.

In WVA regime, i.e., when $|\varphi|/2\ll\delta$, the state of the light and the mirror in Eq. (\ref{a13}) becomes
\begin{eqnarray}
|\psi_{click}^{wva}\rangle&=&\exp[\alpha e^{-ir\omega_{m}t}a_{c}^{\dagger}(\varphi c^{\dagger}-{\varphi}^{\ast}c)/2]  \nonumber \\
&&|\alpha e^{-ir\omega_{m}t}\rangle(|0\rangle_{m}+\varphi/2\delta|1\rangle_{m})  \nonumber \\
&=&\exp[\alpha e^{-ir\omega_{m}t}a_{c}^{\dagger}(\varphi c^{\dagger}-{\varphi}^{\ast}c)/2]  \nonumber \\
&&|\alpha e^{-ir\omega_{m}t}\rangle_{c}|\varphi/2\delta\rangle_{m}, \label{a14}
\end{eqnarray}
and its successful postselection probability is $P_{f}=|\alpha|^{2}\delta^{2}$. Obviously, in WVA regime, the backaction caused by the optomechanical coupling interaction ($|\varphi|$) on the system (the coherent light) has been neglected.

Therefore, in the WVA regime, the expected position displacement on the mirror is gievn by
\begin{equation}
\langle q\rangle_{wva}=\sigma(\varphi+{\varphi}^{\ast})(|\alpha|^{2}/2+1/2\delta).  \label{a16}
\end{equation}
In addition to the classical light effect ($\propto|\alpha|^{2}/2$), this result also shows that, corresponding to the single photon state in WVA regime\cite{Li14,Carrasco19}, the expected mirror's position displacement is $1/2\delta$ times $\sigma(\varphi+{\varphi}^{\ast})$.
\begin{figure}[tbp]
\includegraphics[scale=0.42]{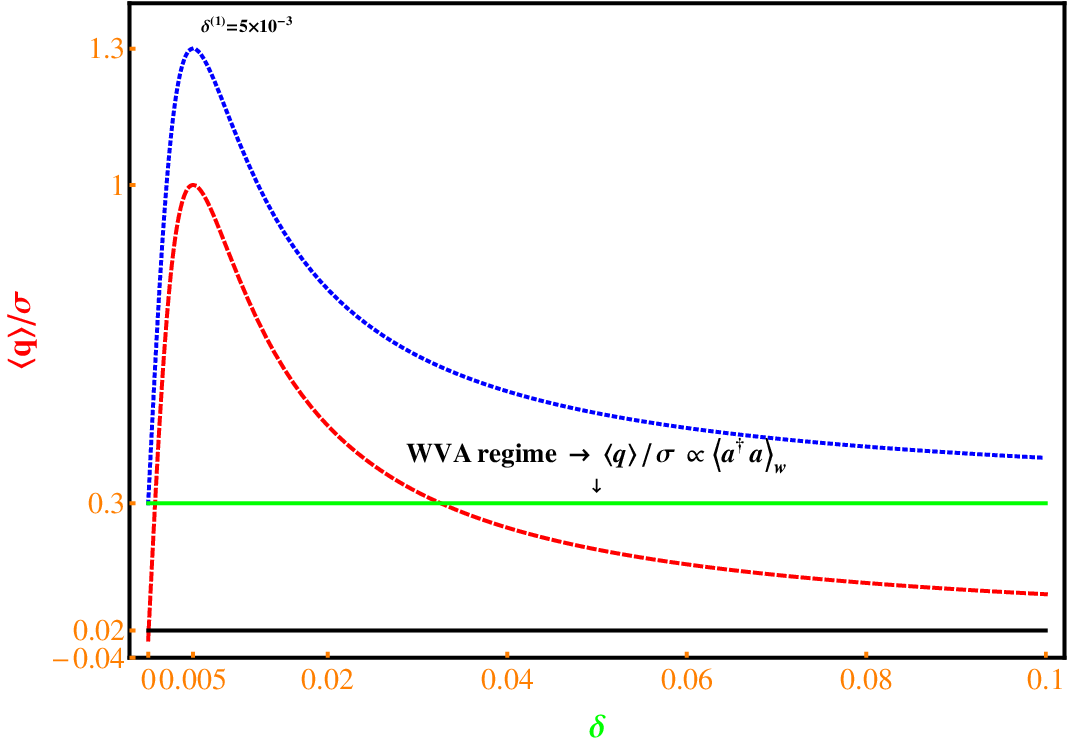}
\caption{The expected position displacement $\langle q\rangle/\sigma$ of the mirror as a function of $\delta$ with $\omega_{m}t=\pi$, $k=0.005$, $|\alpha|^{2}=30$. The mirror's position displacement reaches $\sigma$ outside the WVA regime, when $\delta^{(1)}=\varphi/2$. Note that the success postselection probabilities $P_{f}$ associated to this high amplification factor can always remain high by adjusting $|\alpha|^{2}$, since $P_{f}$ is proportional to $|\alpha|^{2}$.}
\end{figure}

Considering two cases where the photons are detected in the bright port and no photons are detected in the dark, or the photons are detected in the bight port and one photon is detected in the dark port at the same time. If we assume that the interaction time $t$ obeys the relation $\cos\omega_{m}t=-1$, then no photons are found in the optomechanical cavity. In addition, if we further assume that $\omega_{m}t=(2n+1)\pi$, where $n$ is an integer number, then the mirror's position displacement produced by the evolution $U(t)$ of the Hamiltonian in Eq. (\ref{a1}) is maximum. Fig. 2 show that the expected mirror's position displacement $\langle q\rangle/\sigma$ for the click (blue line) and no-click events (green line) versus the postselection parameter $\delta$ with $\omega_{m}t=\pi$, $k=0.005$, $|\alpha|^{2}=30$. Moreover, for weak measurement without the postselection (see appendix C), the expected mirror's position displacement $\langle q\rangle/\sigma=(\varphi+\varphi^{\ast})$ caused by one photon is given (black line). It can be seen clearly from Fig. 2 that when $\delta=\varphi/2=0.005$ the amplification (blue-dotted) can exceed the strong coupling limit $\sigma$ (the level of the vacuum fluctuation) outside the WVA regime, which caused by the classical light effect and the single photon nonlinear effect of weak measurement, and is manifestly always larger than the amplification for the no-click case (green line).

To directly observe the weak measurement amplification of the added one photon in the dark port, in Fig. 2 we also plot
$\langle q\rangle_{click}-\langle q\rangle_{no-click}$ versus the postselection parameter $\delta$ (red line). It is easy to see that this quantity is independent of $|\alpha|^{2}/2$, and the amplified effect of the added one photon is given by
\begin{equation}
\langle q\rangle_{click}-\langle q\rangle_{no-click}=\delta\sigma(\varphi+{\varphi}^{\ast})/(2\delta^{2}+|\varphi|^{2}/2)]  \label{a17}
\end{equation}
Note that outside the WVA regime, the maximum value is $\sigma$ (the strong coupling limit) when $\delta=\varphi/2$, which caused by the single photon nonlinear effect of weak measurement. Therefore, the amplification factor $Q$ is $Q=\pm1/2(\varphi+\varphi^{\ast})$. For example, when $k=0.005$ and $\omega_{m}t=\pi$, $Q$ is $\pm50$. Based on weak measurement, for each $\delta$ and $|\varphi|$, the mean photon number $|\alpha|^{2}$ is adjusted so as to keep the probability of photon detection ($P_{f}=|\alpha|^{2}(\delta^{2}+|\varphi|^{2}/4)$) low, which is the necessary condition for a photon detection to add one photon to the inferred photon number. It can be seen that $P_{f}$ is proportional to the mean photon number $|\alpha|^{2}$. All in all, regardless of the WVA regime or outside the WVA regime, the amplification value is independent of the mean photon number $|\alpha|^{2}$.
\begin{figure}[tbp]
\includegraphics[scale=0.45]{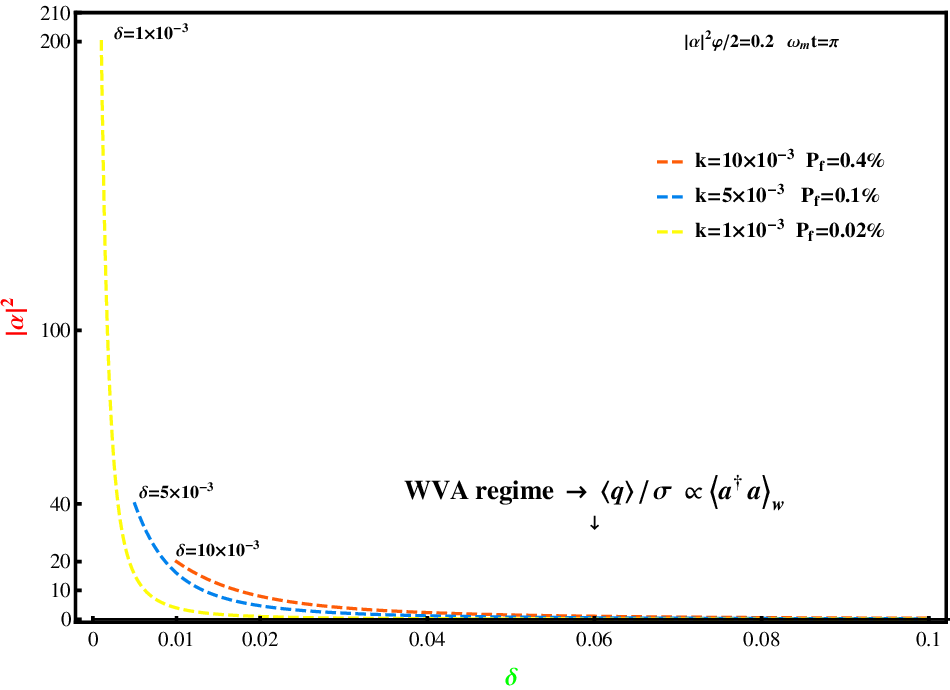}
\caption{The mean photon number $|\alpha|^{2}$ versus the postselection parameter $\delta$ with $\omega_{m}t=\pi$ for the different postselection probability $P_{f}=0.4\%$, $k=0.01$(red line), $P_{f}=0.1\%$, $k=0.005$ (blue line), and $P_{f}=0.02\%$, $k=0.001$ (yellow line).}
\end{figure}

Fig. 3 clearly shows that, when $P_{f}$ is fixed and as $\delta$ becomes smaller, the mean photon number $|\alpha^{2}$ of the coherent light becomes larger. However, outside the WVA regime, if the optomechanical coupling parameter $k$ range is selected appropriately, i.e.,
$0.001\leq k\leq0.01$, and when $\delta$ tends to and is equal to $k$, $P_{f}$ at the strong coupling limit can be improved by $|\alpha|^{2}$, and is in the range $0.02-0.4\%$. In\cite{Li14,Carrasco19} an implementation of WVA with one photon and optomechanical system was proposed as a method for measuring the small coupling constant between a single photon and the mechanical oscillator. In the same condition, $P_{f}$ at the strong coupling limit is in the range $2\times10^{-6}-2\times10^{-4}$. The probability $P_{f}$ of the former is much higher than that of the latter, and is increased by about one to two orders of magnitude. Thefore, outside the WVA regime, weak measurement amplification caused by the coherent light is feasible under current experimental conditions. Obviously, this is an unexpected result. The above result stands as one of the main results of our work, as it breaks the constraint that it is difficult to detect outside the WVA regime.

\section{Weak value amplification in optomechanical system based on a weak coherent light}

\subsection{ Weak value of photon number}
Now we calculate weak value of photon number in an arm of the interferometer versus the postselection parameter
$\delta=(\cos\theta-\sin\theta)/\sqrt{2}$. In Fig. 1, the single photon detector in dark port and the photon detector in bright port are modeled as a beam splitter. The initial state of the weak coherent light can be written in the arm $c$ and the arm $d$ as
\begin{eqnarray}
|\psi_{i}\rangle&=&|\alpha e^{-ir\omega_{m}t}/\sqrt{2}\rangle_{a}|\alpha e^{-ir\omega_{m}t}/\sqrt{2}\rangle_{b}  \nonumber \\
&=&|\alpha e^{-ir\omega_{m}t}(\cos\theta+\sin\theta)/\sqrt{2}\rangle_{c} \nonumber \\
&&|\alpha e^{-ir\omega_{m}t}(\sin\theta -\cos\theta)/\sqrt{2}\rangle_{d} \label{a18}
\end{eqnarray}

After the weak interaction in Eq. (\ref{a1}), the state in the detected arm $d$ is projected onto $|1\rangle_{d}$, in the language of weak measurement the postselected state of the coherent light is
\begin{equation}
|\psi_{f}\rangle=|\alpha e^{-ir\omega_{m}t}(\cos\theta+\sin\theta)/\sqrt{2}\rangle_{c}|1\rangle_{d},  \label{a19}
\end{equation}

The weak value of the photon number in an arm of the interferometer caused by the coherent light is
\begin{eqnarray}
\langle a_{a}^{\dagger}a_{a}\rangle_{w}&=&\langle\psi_{f}|a_{a}^{\dagger}a_{a}|\psi_{i}\rangle/\langle\psi_{f}|\psi_{i}\rangle  \nonumber \\
&\approx&|\alpha|^{2}/2+1/2\delta,  \label{a20}
\end{eqnarray}
where $a_{a}^{\dagger}a_{a}$ can be written in the arm $c$ and the arm $d$ as
\begin{eqnarray}
a_{a}^{\dagger}a_{a}&=&\cos^{2}\theta a_{c}^{\dagger}a_{c}+\cos\theta\sin\theta a_{c}^{\dagger}a_{d}  \nonumber \\
&+&\cos\theta\sin\theta a_{c}a_{d}^{\dagger}+a_{d}^{\dagger}a_{d}+\sin^{2}\theta a_{d}^{\dagger}a_{d}  \label{a21}
\end{eqnarray}
with $\cos\theta=(1-\delta)/\sqrt{2}$ and $\sin\theta=(1+\delta)/\sqrt{2}$. Now if for the single-photon state, the mirror's position displacement without the postselection in the arm $a$ is $\sigma(\varphi+{\varphi}^{\ast})$, the expected position displacement on the mirror caused by the cohernet light will be
\begin{eqnarray}
\langle q\rangle_{wva}&=&\sigma(\varphi+{\varphi}^{\ast})\langle a_{a}^{\dagger}a_{a}\rangle_{w}  \nonumber \\
&=&\sigma(\varphi+{\varphi}^{\ast})(|\alpha|^{2}/2+1/2\delta ), \label{a22}
\end{eqnarray}
which recovers the result of Eq. (\ref{a16}).

To directly observe the WVA effect of the added photon in the dark port, in Fig. 2 we have plotted
$\langle q\rangle_{wva}-\langle q\rangle_{no-click}$ (the WVA effect of the single photon) versus the the postselection parameter
$\delta$\ (red line). It is easy to see that this quantity is independent of $|\alpha|^{2}/2$, and contains only the expected mirror's position displacement of the added photon and its amplified effect is given by
\begin{eqnarray}
\langle q\rangle_{wva}-\langle q\rangle_{no-click}&=&\delta\sigma(\varphi+{\varphi}^{\ast})/2\delta  \nonumber \\
&=&\delta\sigma (\varphi+{\varphi}^{\ast})\langle a_{a}^{\dagger}a_{a}\rangle_{w,1},  \label{a23}
\end{eqnarray}
where $\langle a_{a}^{\dagger}a_{a}\rangle_{w,1}=1/2\delta$ is the weak value of one photon. For example, when $\varphi\sim10^{-3}$ we can set $\delta=0.05$ and displace the mirror as if $10$ photons hit the mirror. The corresponding successful postselection probability $P_{f}$ ($P_{f}=|\alpha|^{2}\delta^{2}$) is $7.5\%$ when $|\alpha|^{2}=30$, with much higher as compared to the successful postselection probability
$P_{f}=2.5\times10^{-3}$ ($P_{f}=\delta^{2}$) of WVA with one photon and optomechanical system\cite{Li14,Carrasco19}. The results show that $P_{f}$ of WVA with the coherent light and optomechanical system can be adjusted by the mean photon number $|\alpha|^{2}$. Table I and Fig. 2 present the weak value and the postselection probability for $\delta\sim10^{-2}$ and $|\alpha|^{2}=30$. In this case, the weak value of one photon goes from $6.25$ to $50$, with the success postselection probabilities in the range $0.3-19.2\%$.
\begin{table}[tbp]
\setlength{\tabcolsep}{6mm}{
\begin{tabular}{lccr}
\hline\hline
$\delta$ & $\langle a^{\dagger}a\rangle_{w,1}$ & $|\alpha|^{2}$ & $P_{f}(\%)$ \\
$0.1$ & 5 & 30 & 30 \\
$0.08$ & 6.25 & 30 & 19.2 \\
$0.06$ & 8.3 & 30 & 10.8 \\
$0.04$ & 12.5 & 30 & 4.8 \\
$0.02$ & 25 & 30 & 1.2 \\
$0.01$ & 50 & 30 & 0.3 \\ \hline\hline
\end{tabular}}
\caption{As the magnitude of the postselection parameter $\delta$ in the dark port decreases, the magnitude of the weak value of the photon
number $\langle a^{\dagger}a\rangle_{w,1}$ is increased whereas the success postselection probability $\delta^{2}|\alpha|^{2}$ decreases rapidly. Notice that WVA (weak values outside the spectrum of $a^{\dagger}a$) begins when $\delta$ is slightly below $0.1$. Recall also that the WVA regime occurs when $|\varphi|/2\ll\delta$.}
\label{table:table1}
\end{table}

When the postselection is successful, i.e., when a single photon is detected in the dark port and the photons are detected in the bright port at the same time, the state of the mirror is given by
\begin{eqnarray}
|\psi_{click}^{m}\rangle&=&P_{f}^{-1}\langle\psi_{f}|\exp[\alpha e^{-ir\omega_{m}t}(a_{c}^{\dagger}+a_{d}^{\dagger})  \nonumber \\
&&(\varphi c^{\dagger}-{\varphi}^{\ast}c)/2]|\psi_{i}\rangle|0\rangle_{m}  \nonumber \\
&=&P_{f}^{-1}\exp[|\alpha|^{2}(\varphi c^{\dagger}-{\varphi}^{\ast}c)/2](\delta\alpha|0\rangle_{m}  \nonumber  \\
&+&\varphi\alpha/2|1\rangle_{m}).  \label{a24}
\end{eqnarray}
The WVA regime applies if $\delta^{2}\gg|\varphi|^{2}/4$, one recovers the weak measurement prediction
\begin{equation}
|\psi_{click}^{m}\rangle=||\alpha|^{2}\varphi+\varphi/2\delta\rangle_{m},  \label{a25-1}
\end{equation}
which is a coherent state with a largely enhanced the mirror's position displacement. Notice that the weak value in Eq. (\ref{a20}) is restricted by $\langle a_{a}^{\dagger}a_{a}\rangle_{w}\ll(\varphi+\varphi^{\ast})^{-1}$ in order to neglect the backaction on the system. On the other hand, if $\delta^{2}\leq|\varphi|^{2}/4$, the postselection is significantly modified by the backaction of the pointer on the system. When $\delta^{2}=|\varphi|^{2}/4$, which corresponds to outside the WVA regime. Fig. 2 clearly shows that the expected mirror's position displacement $\langle q\rangle/\sigma$ is shown by the WVA regime and outside the WVA regime.

When the postselection fails, i.e., when no photon is detected in the dark port and the photons are detected in the bright port, the state of the mirror is given by
\begin{eqnarray}
|\psi_{no-click}^{m}\rangle&=&\langle\psi_{f}|_{\perp}\exp[\alpha e^{-ir\omega_{m}t}(a_{c}^{\dagger}  \nonumber \\
&+&a_{d}^{\dagger})(\varphi c^{\dagger}-{\varphi}^{\ast}c)/2]|\psi_{i}\rangle|0\rangle_{m}  \nonumber\\
&=&||\alpha|^{2}\varphi/2\rangle_{m},  \label{a25-2}
\end{eqnarray}
where $|\psi_{f}\rangle_{\perp}=|\alpha e^{-ir\omega_{m}t}(\cos\theta+\sin\theta)/\sqrt{2}\rangle_{c}|0\rangle_{d}$ is orthogonal to
$|\psi\rangle_{f}$, i.e., $\langle\psi_{f}|_{\perp}|\psi\rangle_{f}=0$. From (\ref{a12}) and Eq. (\ref{a22}), it is clear that the mirror's position displacement of the classic light, without the postselection, and WVA do both not reach the level of vacuum fluctuation $\sigma$. This means that lot of data are required to estimate the centroid of the probability distribution given by the states (\ref{a25-1}) and (\ref{a25-2}), i.e., the quantum-limited SNR is below unity. When the postselection are successful and fails, the mirror's position displacement should be observed\cite{Clerk08,Vanner11}. Under certain conditions, described along this work, the expectation value of the position $q$ with the postselection will correspond to the weak value of the photon number.

\subsection{ Small quantity expansion about time for amplification}

To minimal the influence of environment nosie, we choose the time $\omega_{m}t=\pi$ as the detection time in the follow discussion. When
$k|\alpha|^{2}\ll1$ and at time $\omega_{m}t=\pi$, with observing the maximum amplification effect, we can obtain the approximation of $|\psi_{click}^{m}\rangle$ in Eq. (\ref{a24}),
\begin{equation}
|\psi_{click}^{m}\rangle=(\delta\alpha|0\rangle_{m}+k\alpha|1\rangle_{m})/\sqrt{2}+k^{2}\alpha|\alpha|^{2}|2\rangle_{m}.  \label{a25}
\end{equation}
We note that in the paper by Li et al.\cite{Li14}, they find $|\psi_{click}^{m}\rangle$ proportional to $(|0\rangle_{m}+|1\rangle_{m})/\sqrt{2}$ instead of the above superposition state. Here we use the approximation
$\exp[|\alpha|^{2}(\varphi c^{\dagger}-{\varphi}^{\ast}c)/2]\approx 1+|\alpha|^{2}(\varphi c^{\dagger}-{\varphi}^{\ast}c)/2$ to get the above superposition state. Substituting Eq. (\ref{a25}) into Eq. (\ref{B1}), the expected value of the displacement operator $q$ is given by
\begin{equation}
\langle q\rangle_{click}=(2k\delta+4k^{3}|\alpha|^{2})/(\delta^{2}+k^{2}) \label{a26}
\end{equation}
which gets its maximal value $(1+2k|\alpha|^{2})\sigma$ when $\delta\alpha=k\alpha$, outside the WVA regime.

On the other hand, the fluctuations of the position operator $q$ are given by
\begin{eqnarray}
\Delta q&=&\sqrt{Tr(|\psi_{click}^{m}\rangle\langle\psi_{click}^{m}|q^{2})-Tr(|\psi_{click}^{m}\rangle\langle\psi_{click}^{m}|q)^{2}}  \nonumber \\
&=&\sigma  \label{a27}
\end{eqnarray}

Thus, at time $\omega_{m}t=\pi$ the quantum-limited SNR is given by
\begin{equation}
\langle q\rangle_{click}/\Delta q=(1+2k|\alpha|^{2}).  \label{a28}
\end{equation}
Therefore, outside the WVA regime, the SNR will be increased with weak measurement cause by coherent light. However, to consider weak measurement amplification of the added photon in Eq. (\ref{a17}), its quantum-limited SNR is given by
\begin{equation}
(\langle q\rangle_{click}-\langle q\rangle_{no-click})/\Delta q=1. \label{a29}
\end{equation}
Based on the conclusion of Fig. 3, the maximal amplification value outside the WVA regime can reach the level of the vacuum fluctuation and is
detectable.

\subsection{Dissipation}
For the completeness of the discussion, the damping of the mirror is also considered\cite{Scully17}. Taking into account of dissipation, the master equation of the mechanical system\cite{Bose97} is given by (set $\hbar=1$)
\begin{eqnarray}
d\rho(t)/dt&=&-i[H,\rho(t)]  \nonumber \\
&+&\gamma_{m}[{2c\rho(t)c^{\dagger}-c^{\dagger}c\rho(t)-\rho(t)c^{\dagger}c]/2},  \label{ii}
\end{eqnarray}
where $\gamma_{m}$ is the damping constant.

Similar to the result of the dissipation in Ref.\cite{Li14}, because the actual $\gamma=\gamma_{m}/\omega_{m}$ can be very small
($\gamma =5\times10^{-7}$ in Proposed device no. 2\cite{Bouwmeester12}) and is almost close to $\gamma=0$, all the amplification values in the presence of the damping are almost unscathed.

\section{Discussion and Conclusion}
We now present some discussions on the experimental parameters for implementation of the postselection weak-coupling optomechanical model. In
principle, the studies in this work are general, and it can be implemented with various optomechanical systems. Nevertheless, we should point out that some used parameters are accessible with current experiments, but there still exists some challenge for current experimental technology. To observe the expected mirror's displacement outside the WVA regime, the maximal amplification value generation time
$t=\pi/\omega_{m}$ is required to be shorter than the lifetime $1/\kappa$ of the cavity photon, with decay rate $\kappa$ of optical cavity, which leads to the resolved-sideband condition $\omega_{m}\gg\kappa$. In the experiment requirements of our scheme, we used the following parameters: $k=g/\omega_{m}\approx0.001-0.01$, $\kappa/\omega_{m}\approx0.01-0.5$, $\gamma_{m}/\omega_{m}\approx10^{-7}-10^{-6}$. For example, in optomechanical device\cite{Bouwmeester12}, when the scaled coupling parameter is $k=g/\omega_{m}=0.005$, the sideband-resolution measure $\omega_{m}/\kappa=3$, and the mechanical frequency is $f_{m}=4.5$ kHz ($\omega_{m}=9\pi$ kHz), thus the optomechanical coupling strength is $g=141.3$ Hz and the decay rate of the cavity is $\kappa=3\pi$ kHz, which are accessible with the current experimental conditions. In the WVA regime, the amplification factor $Q$ correspond to the weak value of the coherent light
$\langle a_{a}^{\dagger}a_{a}\rangle_{w}=|\alpha|^{2}/2+1/2\delta$ and the weak value of one photon
$\langle a_{a}^{\dagger}a_{a}\rangle_{w,1}=1/2\delta$, respectively. As long as $k\ll\delta$ and $k|\alpha|^{2}\ll1$, and combined with the above conditional parameters in\cite{Bouwmeester12}, it can be achieved under the current experimental conditions. In our scheme, the successful postselection probability $P_{f}$ is increased outside the WVA regime, it becomes to observe the amplification effect, as compared to the previous schemes\cite{Li14,Carrasco19}. Simultaneously measuring the position and momentum of the mirror is impossible due to the quantum uncertainty principle. However, it is possible to measure only one quadrature component of the mechanical motion, such as the position, to an arbitrary precision\cite{Clerk08}. This idea can be used for a full reconstruction of the mechanical quantum state, extracting its Wigner density using quantum state tomography\cite{Vanner11}.

In summary, we propose to use weak coherent light to produce the amplification effect of one photon in the postselection weak-coupling
optomechanical system. We focus on weak measurement amplification in the WVA regime and outside the WVA regime. In the WVA regime
($\delta\gg|\varphi|$), when $\delta$ is small, the amplification factor $Q$ of one photon and the coherent light is $|\alpha|^{2}/2+1/2\delta $, and the amplification factor $Q$ of the coherent light without postselection is $|\alpha|^{2}/2$. Then the amplification factor $Q$ of one photon is $1/2\delta$. The result show that while the differential position displacement between the successful and failed postselection grows as $1/\delta $, the size of the postselected data set is determined by the successful postselection probability $|\alpha|^{2}\delta^{2}$. As an example, if $\delta=0.05$, then the weak value of one photon and the coherent light is approximately equal to $25$ when $|\alpha|^{2}=30$, and the weak value of one photon is approximately equal to $10$, and the associated probability of postselection is nearly $7.5\%$. However, the successful postselection probability for the single photon case is $0.25\%$\cite{Li14,Carrasco19}. The probability of the former is 30 times that of the latter. Hence, in the case of weak coherent light, we were able to maintain a substantial set of postselections even for small $\delta$ simply by adjusting $|\alpha|^{2}$ accordingly. This case might be useful for the estimation of $g$, when the mirror is operated at the quantum level. In particular, in Fig. 3, outside the WVA regime, if the optomechanical coupling parameter $k$ range is selected appropriately, i.e., $0.001\leq k\leq0.01$, and when $\delta$ is equal to $\varphi/2$ with $\omega_{m}t=\pi$, $P_{f}=|\alpha|^{2}(\delta^{2}+|\varphi|^{2}/4)$ at the strong coupling limit can be adjusted by $|\alpha|^{2}$ to increase, as compared to weak measurement schemes with one photon and optomechanical system\cite{Li14,Carrasco19}. Becuase $P_{f}$ becomes higher outside the WVA regime, it is beneficial to observe the maximum magnification value under the current experimental conditions. Moreover, the quantum-limited SNR is increasing, even reaching $1$. It is essential to note that the WVA technique in this scheme is advantageous only if one considers the number of trials (and not the number of photons) as the measurement resource.

\section*{Acknowledgement}
This work was supported by the National Natural Science Foundation of China under Grant 11947100, Key Project of Chinese National Programs for
Fundamental Research and Development (973 Program) under Grant 2017YFA0303400, the China Postdoctoral Science Foundation Funded Project
under Grant 2020M680317, the Natural Science Foundation of Shaanxi Province under Grant 2021JM-414, the Education department Program of Shaanxi under Grant 20JK0982, and the Doctoral Scientific Research Foundation of Yan'an University under Grant YDBK2016-04.

\appendix

\section{Simplification of the combined state $|\psi_{evo}\rangle$ after the second imbalanced beam splitter}
In the representation of the arm $c$ and the arm $d$, the state $|\psi_{evo}\rangle$ (\ref{a5}) in the main text is given by
\begin{eqnarray}
|\psi_{evo}\rangle&=&\exp[-ir\omega_{m}t(a_{c}^{\dagger}a_{c}+a_{d}^{\dagger}a_{d})]\exp[i\phi(\cos^{2}\theta a_{c}^{\dagger}a_{c} \nonumber \\
&+&\cos\theta\sin\theta(a_{c}^{\dagger}a_{d}+a_{c}a_{d}^{\dagger})+\sin^{2}\theta a_{d}^{\dagger}a_{d})^{2}]  \nonumber \\
&&\exp[(\cos^{2}\theta a_{c}^{\dagger}a_{c}+\cos\theta\sin\theta(a_{c}^{\dagger}a_{d}+a_{c}a_{d}^{\dagger})  \nonumber \\
&+&\sin^{2}\theta a_{d}^{\dagger}a_{d})(\varphi c^{\dagger}-{\varphi}^{\ast}c)]\exp[-i\omega_{m}tc^{\dagger}c]|\alpha \nonumber \\
&&(\cos\theta+\sin\theta)/\sqrt{2}\rangle_{c}|\alpha(\sin\theta   \nonumber \\
&-&\cos\theta)/\sqrt{2}\rangle_{d}|0\rangle_{m}.  \label{A1}
\end{eqnarray}
The second imbalanced beam splitter in the main text implies that $\theta\approx\pi/4$. For $\theta=\pi/4+\delta$ ($\delta\ll1$, called
the postselection parameter), we expand $\cos\theta$ and $\sin\theta$ as $\cos\theta=(1-\delta)/\sqrt{2}$ and
$\sin\theta=(1+\delta)/\sqrt{2}$, respectively. Because coherent light is weak and the optomechanical coupling interaction is also weak, i.e., $|\varphi|=k\sqrt{2}(1-\cos\omega_{m}t)\ll1$ and $k\ll1$, so $\exp[k^{2}X]\approx1$ and $\exp[\delta kY]\approx1$, where $X$ and $Y$ are both polynomials. Then the state in Eq. (\ref{A1}) becomes
\begin{eqnarray}
|\psi_{evo}\rangle&=&\exp[-ir\omega_{m}t(a_{c}^{\dagger}a_{c}+a_{d}^{\dagger}a_{d})]\exp[(a_{c}^{\dagger}a_{c}  \nonumber \\
&+&a_{d}^{\dagger}a_{d})(\varphi c^{\dagger}-{\varphi}^{\ast}c)/2]\exp[(a_{c}^{\dagger}a_{d}  \nonumber \\
&+&a_{c}a_{d}^{\dagger})(\varphi c^{\dagger}-{\varphi}^{\ast}c)/2]|\alpha\rangle_{c}|\delta\alpha\rangle_{d}|0\rangle_{m}.  \label{A2}
\end{eqnarray}

For the sake of making the analysis simple, in the phase space we define the momentum operator
$p_{new}=i(\varphi c^{\dagger}-{\varphi}^{\ast}c)/(2|\varphi|\sigma)$ and the position operator
$q_{new}=\sigma (\varphi c^{\dagger}-{\varphi}^{\ast}c)/|\varphi|$ with $[q_{new},p_{new}]=i$. Apply the operator $e^{W}$ to perform unitary transformation on two boson operators $a_{c}$ ($a_{d}$) and $a_{c}^{\dag }$ ($a_{d}^{\dagger}$), where
$W=|\varphi|q_{new}(a_{c}^{\dagger}a_{d}+a_{c}a_{d}^{\dagger})/2\sigma$. From the Baker-Hausdorff expansion\cite{Scully17}
\begin{eqnarray}
e^{A}Be^{-A} &=&B+[A,B]+[A,[A,B]]/2!  \nonumber \\
&+&[A,[A,[A,B]]]/3!+\cdots,  \label{A3}
\end{eqnarray}
we can see
\begin{eqnarray}
e^{W}a_{c}(a_{d})e^{-W}&=&a_{c}(a_{d})\cosh(|\varphi|q_{new}/2\sigma) \nonumber \\
&-&a_{d}(a_{c})\sinh(|\varphi|q_{new}/2\sigma)  \label{A4-1}
\end{eqnarray}
and
\begin{eqnarray}
e^{W}a_{c}^{\dagger}(a_{d}^{\dagger})e^{-W}&=&a_{c}^{\dagger}(a_{d}^{\dagger})\cosh (|\varphi|q_{new}/2\sigma )  \nonumber \\
&+&a_{d}^{\dagger}(a_{c}^{\dagger})\sinh (|\varphi|q_{new}/2\sigma ). \label{A4-2}
\end{eqnarray}
In order to obtain the above result, here we use two equations
\begin{equation}
[W,a_{c}(a_{d})]=-|\varphi|q_{new}/2\sigma a_{d}(a_{c})  \label{A5-1}
\end{equation}
and
\begin{equation}
[W,a_{c}^{\dagger}(a_{d}^{\dagger})]=|\varphi|q_{new}/2\sigma a_{d}^{\dagger}(a_{c}^{\dagger})  \label{A5-2}
\end{equation}

Using the fact that
\begin{equation}
e^{W}f(\{X_{i}\})e^{-W}=f(\{e^{W}X_{i}e^{-W}\})  \label{A6}
\end{equation}
for any function $f$, unitary operator $e^{W}$, and arbitrary set of operators $\{X_{i}\}$, and using $e^{x(a^{\dagger}a)}ae^{-x(a^{\dagger}a)}=ae^{-x}$ and $e^{x(a^{\dag }a)}a^{\dagger}e^{-x(a^{\dagger}a)}]=a^{\dagger}e^{x}$, then the state in Eq. (\ref{A2}) can be simplified to
\begin{eqnarray}
|\psi_{evo}\rangle&=&\exp[Ja_{c}^{\dagger}-J^{\ast}a_{c}]\exp[Fa_{d}^{\dagger}-F^{\ast}a_{d}]|0\rangle_{c}|0\rangle_{d}|0\rangle_{m},
\nonumber \\
&\approx&\exp[\alpha e^{-ir\omega_{m}t}(a_{c}^{\dagger}+a_{d}^{\dagger})(\varphi c^{\dagger}-{\varphi}^{\ast}c)/2]  \nonumber \\
&&|\alpha e^{-ir\omega_{m}t}\rangle_{c}|\delta\alpha e^{-ir\omega_{m}t}\rangle_{d}|0\rangle_{m}  \label{A7}
\end{eqnarray}
where
\begin{eqnarray}
J&=&e^{-ir\omega_{m}t}\alpha[\exp(|\varphi|q_{new}/\sigma)+1
\nonumber \\
&+&\delta\exp(|\varphi|q_{new}/\sigma)-\delta]/2  \label{A8}
\end{eqnarray}
and
\begin{eqnarray}
F&=&e^{-ir\omega_{m}t}\alpha[\exp(|\varphi|q_{new}/\sigma)-1
\nonumber \\
&+&\delta\exp(|\varphi|q_{new}/\sigma)+\delta]/2.  \label{A9}
\end{eqnarray}
This Eq. (\ref{A7}) is equation (\ref{a9}) in the main text. In order to obtain the result of Eq. (\ref{A7}), here we use the equation
\begin{equation}
e^{A+B}=e^{A}e^{B}e^{-[A,B]/2}=e^{B}e^{A}e^{[A,B]/2},  \label{A10}
\end{equation}
where $[A,[A,B]]=[B,[A,B]]=0$ and the approximation
\begin{equation}
\exp[(\varphi c^{\dagger}-{\varphi}^{\ast}c)/2]\approx1+(\varphi c^{\dagger}-{\varphi}^{\ast}c)/2.  \label{A11}
\end{equation}

\section{The expected displacement of the pointer observable $M$}
The expected displacement of the pointer observable $M$ ($M=q,p$), or the average displacement of the pointer observable $M$, is given by
\begin{equation}
\langle M\rangle=Tr(\rho_{f}M)/Tr(\rho_{f})-Tr(\rho_{i}M), \label{B1}
\end{equation}
where $\rho_{f}$ is the postselected pointer state based on weak mesurement and $\rho_{i}$ is the initial state of the pointer.

\section{The amplification without postselection in optomechanics}

The time evolution operator of the Hamiltonian (\ref{a1}) in the main text is given by
\begin{eqnarray}
U(t)&=&\exp[-ir\omega_{m}t(a_{a}^{\dagger}a_{a}+a_{b}^{\dagger}a_{b})]\exp[i\phi(a_{a}^{\dagger}a_{a})^{2}]  \nonumber \\
&\times&\exp[a_{a}^{\dagger}a_{a}(\varphi c^{\dagger}-{\varphi}^{\ast}c)]\exp[-i\omega_{m}tc^{\dagger}c],  \label{C1}
\end{eqnarray}
where $r=\omega_{0}/\omega_{m}$, $\phi=k^{2}(\omega_{m}t-\sin\omega_{m}t)$, $\varphi=k(1-e^{-i\omega_{m}t})$, $k=g/\omega_{m}$ is the scaled
coupling parameter.

As shown Fig. 1 in main text, we use only single cavity A. When the ground state $|0\rangle_{m}$ is considered as a pointer in cavity A, and if one photon is weakly coupled with the mirror using (\ref{C1}), it can be found that the mirror will be changed from $|0\rangle_{m}$ to a displacement state $|\varphi\rangle_{m}$. Substituting $|\varphi\rangle_{m}$ and $|0\rangle_{m}$ into (\ref{B1}), the expected position displacement on the mirror without postselection will be
\begin{eqnarray}
\langle q\rangle&=&\sigma(\varphi+{\varphi}^{\ast})=2k(1-\cos\omega_{m}t)\sigma.  \label{C2}
\end{eqnarray}

Fom Eq. (\ref{C2}), it can be seen that the position displacement of the mirror caused by radiation pressure of one photon can not more than
$4k\sigma$ for any time $t$. In the literature\cite{Marshall03}, we know that if the displacement of the mirror can be detected experimentally it should be not smaller than $\sigma$, implying that the displacement of the mirror reach strong-coupling limit, so $k=g/\omega_{m}$ can not be bigger than $0.25$ in weak coupling condition\cite{Marshall03}. When $k=g/\omega_{m}\leq 0.25$ in weak-coupling regime, the maximal displacement of the mirror $4k\sigma$ can not be more than $\sigma$, i.e., zero-point fluctuation of the mirror, therefore the displacement of the mirror caused by one photon can not be detected.

\end{document}